\documentclass[smallextended]{svjour3}                    
\smartqed  
\usepackage{graphicx}
\usepackage{amsmath,amsfonts,amscd,amssymb,epsfig}
\usepackage{color}
\journalname{General Relativity and Gravitation}
\newcommand{\be}{\begin{equation}}
\newcommand{\beq}{\begin{equation}}
\newcommand{\eeq}{\end{equation}}
\newcommand{\ee}{\end{equation}}
\newcommand{\bea}{\begin{eqnarray}}
\newcommand{\eea}{\end{eqnarray}}
\newcommand{\ba}{\begin{array}}
\newcommand{\ea}{\end{array}}

\newcommand{\ga}{\gamma}
\newcommand\IZ{\mathbb{Z}}
\newcommand\IR{\mathbb{R}}

\newcommand\IT{\mathbb{T}}
\newcommand{\II}{\mathbb{I}}

\def\rref#1{(\ref{#1})}
\begin{document}

\title{Quantum geometry from $2+1$ AdS quantum gravity on the torus}


\author{J. E. Nelson \and R. F. Picken}


\institute{J. E. Nelson (corresponding author) \at Dipartimento di Fisica Teorica, Universit\`a
degli Studi di Torino\\ 
and\\ 
INFN, Sezione di Torino\\ 
via Pietro Giuria 1, 10125 Torino, Italy\\
  \email{nelson@to.infn.it}
\and  R. F. Picken \at Departamento de Matem\'{a}tica\\ 
and\\ 
CAMGSD - Centro de An\'{a}lise Matem\'{a}tica, Geometria e Sistemas Din\^{a}micos\\
Instituto Superior T\'{e}cnico, Technical University of Lisbon\\
Avenida Rovisco Pais, 1049-001 Lisboa, Portugal\\
\email{rpicken@math.ist.utl.pt}
}

\date{Received: date / Accepted: date}

\maketitle

\begin{abstract}

Wilson observables for $2+1$ quantum gravity with negative cosmological constant, when 
the spatial manifold is a torus, exhibit several novel features: signed area phases relate the observables assigned to homotopic loops, and 
their commutators describe loop intersections, with properties that are not yet fully understood. We describe progress in our 
study of this bracket, which can be interpreted as a $q$-deformed Goldman bracket, and provide a geometrical interpretation in terms of a quantum version of Pick's formula for the area of a polygon with integer vertices.  

\end{abstract}

\section{Introduction} \label{intro} \noindent Interest in the 
construction of a 
commutator for Wilson loops associated to intersecting loops 
on surfaces arose from the study of quantum gravity for spacetimes of 
dimension $2+1$ with negative cosmological constant \cite{NR1,NRZ}. 
When the spatial manifold is a torus, piecewise linear (PL) paths in 
the covering space 
${\IR}^2$ were used to represent loops on the torus, and a quantum 
connection with noncommuting components was introduced to construct quantum 
holonomies whose traces quantize the Poisson bracket structure 
of $2+1$ gravity \cite{NP1,qmp,goldman}. This may be regarded as a 
quantum version of the Goldman bracket (see Section \ref{sec3})
for loops on a surface. The purpose of this paper is to describe 
progress in the study of this bracket

The classical action of $2+1$ gravity with negative cosmological 
constant $\Lambda$ is related to Chern-Simons theory for the gauge 
group $SL(2,\IR)\times SL(2,\IR)$ \cite{wit}. The corresponding classical phase 
space is the moduli space of flat $SL(2,\IR)\times SL(2,\IR)$ 
connections on the spatial manifold, which here is the torus 
$\IT^2=\IR^2/\IZ^2$. This moduli space represents the fundamental 
group of the torus, and has two generators, or generating cycles, 
say $\gamma_1,\, \gamma_2$ which satisfy the single relation 
\be
\ga_1^{\vphantom{-1}}\cdot\ga_2^{\vphantom{-1}}\cdot\ga_1^{-1}\cdot\ga_2^{-1}
  = {\II}.
\label{gp}
\ee
i.e. the generators $\gamma_1,\, \gamma_2$ commute.  The moduli space 
therefore consists of pairs of commuting $SL(2,\IR)$ matrices (up to 
simultaneous conjugation by 
an element of $SL(2,\IR)$), representing the holonomies of the flat 
connection along the two cycles of the torus (up to gauge 
transformation). It was studied in \cite{mod}.

The diagonal, hyperbolic--hyperbolic sector of this moduli space is 
the relevant sector for $2+1$ gravity \cite{Ezawa1,Ezawa2}, and each 
sector can be parametrised by constant connections 
\cite{goldman,mikpic} on the torus of the form
\be
A= (r_1 dx + r_2 dy)
\left(\begin{array}{clcr} 1& 0 \\0& -1 
\end{array}\right)   
\label{conn}
\ee
where $x,\,y$ are coordinates on the torus (with periodicity $1$), or 
on its covering space $\IR^2$. The connections are constant in that 
$r_1,\,r_2$ are global constants (independent of $x$ and $y$). After integration, 
this yields the holonomies:
\be
U_i = \exp \int_{\ga_i} A = \left(\begin{array}{clcr}e^{{r_i}}& 0 \\0& e^{-
{r_i}}\end{array}\right) \quad i=1,2.
\label{hol1}
\ee
The traces of the holonomies \rref{hol1} are precisely the variables used in \cite{NR1} 
to quantize this model of $2+1$ gravity. The relation between them  and the ADM variables 
was described in \cite{CN}. Therefore our present approach can be related to standard descriptions, but a richer picture is obtained by including more general Wilson loops, derived from the quantum connection \rref{conn}.

In Chern-Simons theory components of the connection $A$ (equation \rref{conn}) satisfy
non-trivial Poisson brackets. For the global parameters 
$r_i,\,i=1,2$ this implies the Poisson bracket
\be
\left\{ r_1, r_2\right\} = -\frac{\sqrt {-\Lambda}}{4},
\ee
which is quantized by the commutator
\begin{equation}
[\hat{r}_1, \hat{r}_2] = \hat{r}_1\hat{r}_2 - \hat{r}_2\hat{r}_1 = \frac{i\hbar \sqrt{-\Lambda}}{4}.
\label{rcommrel}
\end{equation}

Further, the commutator \rref{rcommrel} implies that the {(now \it 
quantum matrices}) $\hat U_1, \hat 
U_2$ \rref{hol1} must satisfy {\it by both matrix and operator multiplication}, the 
$q$--commutation relation
\beq
\hat U_1 \hat U_2 = q \hat U_2 \hat U_1 
\label{fund2}
\eeq
where the $q$ parameter\footnote{Note that in $q$ the exponent of 
$\exp$ is dimensionless, when all physical constants are taken into account.} is $q=\exp (- \frac {i \hbar \sqrt{-\Lambda}}{4})$ i.e. the matrices $\hat U_1, \hat U_2$ form a matrix--valued Weyl pair. Equation \rref{fund2} can be understood as a deformation of equation \rref{gp}. Analogous commutators for a general class of Wilson loops give rise to the quantum Goldman bracket, our central theme.

The plan of the paper is as follows. In Section \ref{sec2} some interesting features of the emerging quantum geometry are reviewed. These are the representation of loops on the torus by piecewise--linear (PL) paths between integer points in $\IR^2$, constant matrix--valued connections applied to a much larger class of loops, and a definition of a $q$--deformed representation of the fundamental group where signed area phases relate the quantum matrices assigned to homotopic loops. In Section \ref{sec3} intersections and reroutings, and two quantizations of the Poisson bracket between paths are described: the `direct' quantization and the `refined' quantization. In Section \ref{sec4} we describe new results which represent progress in understanding the quantum nature of intersections (expressed as commutators) of straight paths (i.e. straight in $\IR^2$) and `crooked' paths (those resulting from previous reroutings) by using the concepts of integer points and relative phases for a crooked rerouting.  As a consequence we obtain a quantum version of a formula for the area of a polygon with integer 
vertices. We also show the equivalence of the refined and direct quantization of the bracket for straight paths, and the antisymmetry of the refined bracket.

\section{Piecewise linear paths and quantum holonomy matrices}\label{sec2}
\subsection{Piecewise linear paths}\label{sub21}
\begin{figure}
\begin{center}
\includegraphics[width=16pc]{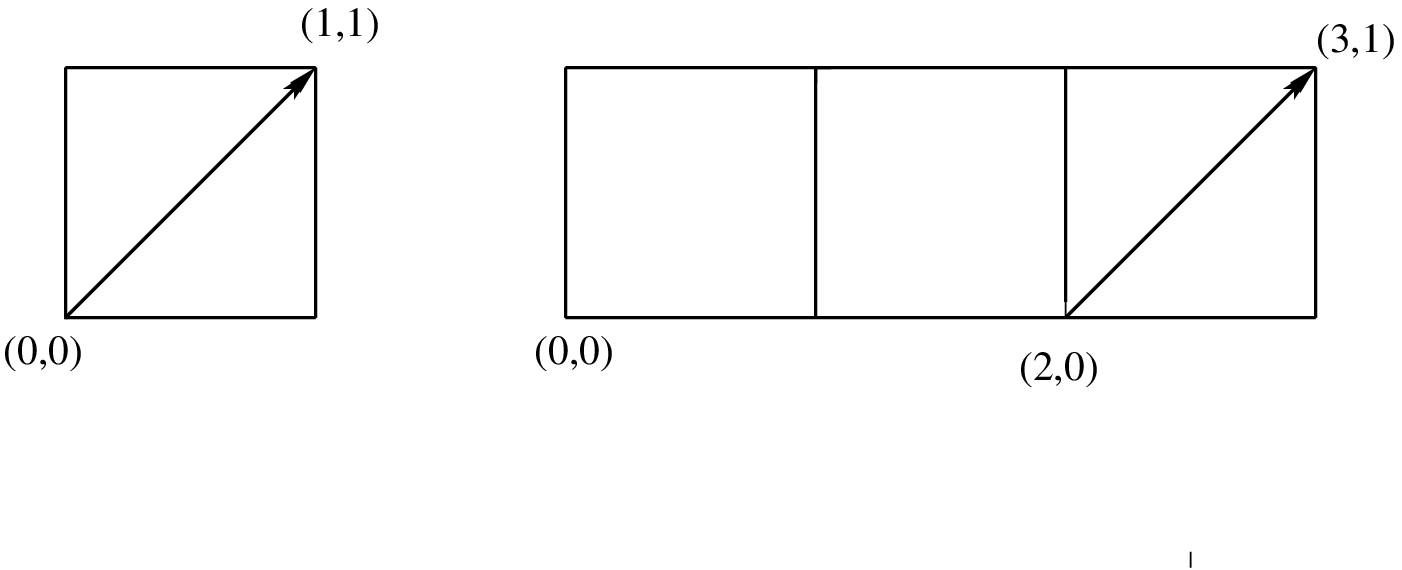}
\end{center}
\caption{\label{fund} The path from $O$ = $(0,0)$ to $(1,1)$ and the path from $(2,0)$ to $(3,1)$ represent the same loop on the torus}
\end{figure}

Loops (closed paths) on the torus $\IT^2=\IR^2/\IZ^2$ can be conveniently identified with paths on its covering space ${\IR}^2$, i.e. we represent all loops on the torus by piecewise-linear (PL) paths on ${\IR}^2$ i.e. the $(x,y)$ plane, between integer points $(m,n)\in \IZ^2$. All these integer points are identified, and correspond to the same point on the torus. It follows that a path in ${\IR}^2$ representing a loop on the torus can be replaced by any parallel path starting at a different integer point, e.g. the path from $O=(0,0)$ to $(1,1)$ represents the same loop as the path from $(2,0)$ to $(3,1)$, as shown in Figure \ref{fund}.

A natural subclass of paths in ${\IR}^2$ are those straight paths denoted $p=(m,n)$ that start at the origin $O =(0,0)$ and end at an integer point $(m,n)\in \IZ^2$. They generalise the cycles $\gamma_1,\, \gamma_2$ (corresponding to the paths $(1,0)$ and $(0,1)$ respectively). When the path $(m,n)$ is a multiple of another integer path, i.e. $(m,n)=c(m',n')$, where $m,n,c,m',n'$ are all integers, with $c\geq 2$, we say it is {\it reducible}. Otherwise it is irreducible. 

The identification between loops on the torus $\IT^2=\IR^2/\IZ^2$ and PL paths in $\IR^2$ can be  further understood by considering the concept of fundamental reduction introduced in \cite{goldman}. There we introduced this concept in order to better study the intersections between paths $p_1$ and $p_2$. It consists of reducing one or more paths to a fundamental domain of 
$\IR^2$, namely the unit square with vertices $(0,0), (1,0)$ and $(1,1), (0,1)$. This works as follows: a path that passes through more than one cell (a unit square in $\IR^2$) consists of ordered segments, each of which passes through only one cell. The fundamental reduction is obtained by superimposing each of these cells, in the order of the segments, on the fundamental domain (the square, or cell, with vertices $(0,0), (1,0)$ and $(1,1), (0,1)$). In this fundamental domain the left and right edges should be identified, and similarly for the top and bottom edges. 

\begin{figure}[hbtp]
\centering
\includegraphics[height=2cm]{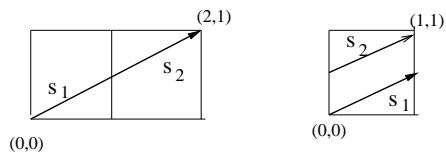}
\caption{The path $(2,1)$ and its fundamental reduction}
\label{21}
\end{figure}

\begin{figure}[hbtp]
\centering
\includegraphics[height=3cm]{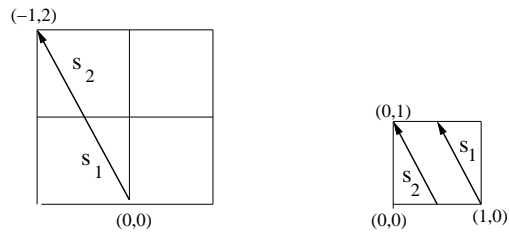}
\caption{The path $(-1,2)$ and its fundamental reduction}
\label{-12}
\end{figure}

Two examples of fundamental reduction for straight paths are shown in Figures \ref{21} and \ref{-12}. Figure \ref{21} shows a straight path in the first quadrant, namely the path $(2,1)$, with its two segments labelled $s_1$ and $s_2$ (in that order), and its reduction to the fundamental domain

\noindent whereas Figure \ref{-12} shows a straight path in the second quadrant, namely $(-1,2)$, and its two segments $s_1$ and $s_2$ (in that order). Note that, when fundamentally reduced, this path starts at $(1,0)$ (not the origin $O=(0,0)$) and ends at $(0,1)$. This is because its first cell does not coincide with the fundamental domain, and is true for non--reduced paths in all quadrants except the first. 
  
It is important that in the fundamental domain diagrams (the diagrams on the right) of Figures \ref{21} and \ref{-12} the left and right edges should be identified, and similarly for the top and bottom edges, i.e. these diagrams depict a loop on the torus.

\subsection{Quantum holonomy matrices and signed area}\label{sub22}
Using the connection \rref{conn} a quantum matrix is assigned to any straight path $(m.n)$ by
\be
\hat{U}_{(m,n)}= \exp \int_{(m,n)} \hat{A} = 
\left(\begin{array}{ll} e^{m\hat{r}_1+ n\hat{r}_2} & 0 
\\ 0 & e^{-m\hat{r}_1 -n\hat{r}_2}
\end{array}\right).
\label{Umn}
\ee
Clearly the assignment \rref{Umn} extends straightforwardly to any PL path between integer points by assigning a quantum matrix to each linear segment of the path, as in \rref{Umn}, and multiplying the matrices in the same order as the segments along the path. This prescription obviously coincides with the general relation:
\be
p\mapsto \hat{U}_p = {\cal P }\exp \int_{p} \hat{A}.
\label{hol}
\ee
where $\cal P$ denotes path-ordering.

\begin{figure}
\begin{center}
\includegraphics[width=16pc]{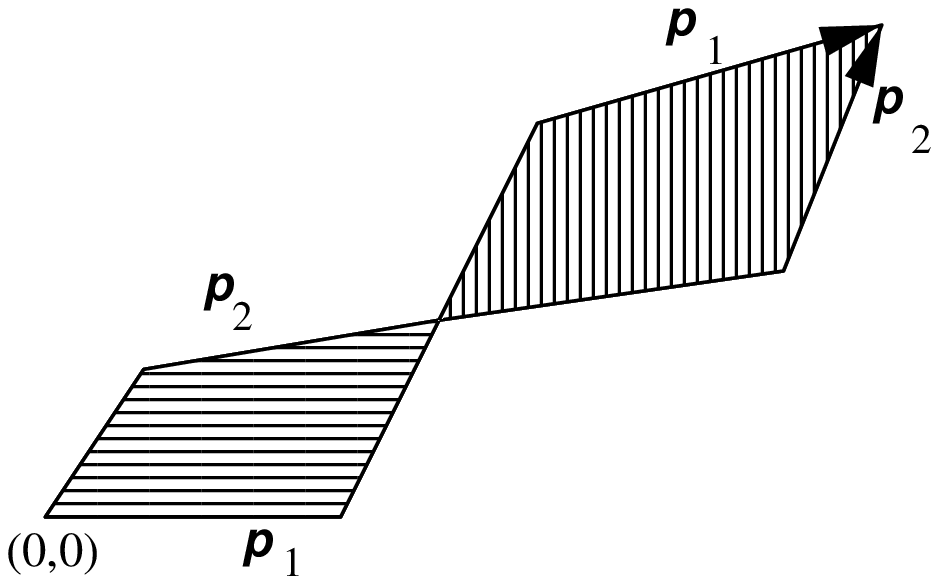}
\end{center}
\caption{\label{signedarea} The signed area between $p_1$ and $p_2$ is the area of the the horizontally shaded region minus the area of the vertically shaded region}
\end{figure}

In the covering space ${\IR}^2$, two homotopic loops on the torus are represented by two PL paths on the plane, $p_1,\, p_2$, with the same integer starting point and the same integer endpoint.  It was shown in \cite{goldman} that the following relationship holds for the respective quantum matrices:
\begin{equation}
\hat{U}_{p_1}=q^{S(p_1,p_2)}\hat{U}_{p_2},
\label{areaphase}
\end{equation}
where $S(p_1,p_2)$ denotes the signed area enclosed between the paths
$p_1$ and $p_2$. Equation \rref {areaphase} generalises equation \rref{fund2}, for which $S(p_1,p_2)$ is the area of the square with vertices $(0,0), \, (0,1), \, (1,0)$ and $(1,1)$, i.e. 1. For the general case, the signed area between two PL paths is defined as follows: for any finite region $R$ enclosed by $p_1$ and $p_2$, if the boundary of $R$ consists of oriented segments of $p_1$ and $p_2^{-1}$ (the path $p_2$ followed in the opposite direction), and is globally oriented in the positive (anticlockwise), or negative (clockwise) sense, this gives a contribution of $+{\rm area}(R)$ , or $-{\rm area}(R)$ respectively, to the signed sum $S(p_1,p_2)$ (otherwise the contribution is zero). See Figure \ref{signedarea}.

In \cite{goldman} this relation between the holonomy matrices for homotopic loops/ paths was interpreted as a $q$-deformed surface group representation, i.e. a deformation of a representation of the fundamental group in terms of matrices. There we used the fact that signed area between paths has the following two properties:
\begin{equation}
 S(p_1,p_3) =  S(p_1,p_2) + S(p_2,p_3), \quad S(p_1p_2,p_3p_4) = S(p_1,p_3) + S(p_2,p_4)
\label{signedareaprops}
\end{equation}
where $p_1p_2$ denotes the concatenation of paths $p_1$ and $p_2$.

Another perspective on this comes from \cite{FMP}, where the concept of $2$-dimensional holonomy was explored. In the present context, $2$-dimensional holonomy may be described as the assignment of an element of a group $G$ to each of a pair of homotopic loops, and simultaneously the assignment of an element of a group $E$ to the homotopy between the two loops, subject to certain consistency relations. Amongst these relations, there is the requirement that the elements assigned to the homotopies behave well under vertical and horizontal composition, which translates precisely to the properties of signed area as in \rref{signedareaprops}. We note also that a crucial ingredient in the construction of \cite{FMP} is a connection $2$-form with values in the Lie algebra of $E$. Indeed, in the present context there is a natural $2$-form, namely the non-vanishing curvature of the connection \rref{conn} (non-vanishing because of the non-commutativity of the components $r_1$ and $r_2$, so that $A\wedge A$ is non-zero, as pointed out in \cite{goldman}). Thus, there are strong hints that the quantum holonomies of $2+1$ gravity may be interpreted in terms of $2$-dimensional holonomy.

\section{Intersecting paths}\label{sec3}
That Wilson loops associated to intersecting paths on surfaces have non--zero Poisson brackets was noted in \cite{gol}. These brackets and the corresponding commutators, for $2+1$--dimensional AdS gravity, were studied in \cite{NRZ}. In all cases paths which are {\it rerouted} at the intersection points appear on the r.h.s. In this Section we review some results about intersections and reroutings, and their classification by integer points. We also show how these concepts appear in the Poisson brackets of Wilson loops, and give some explicit examples.

\subsection{Intersections and reroutings}\label{sub31}
The intersection number between two paths $p_1$ and $p_2$, corresponding to loops on the torus, at an intersection point $S$, is defined to be (for single, not multiple, intersections) $+1$ if the angle between the tangent vector of $p_1$ at $S$ and the tangent vector of $p_2$ at $S$ is between $0$ and $180$ degrees, and $-1$ if it is between $180$ and $360$ degrees.

\begin{figure}[hbpt]
\centering
\includegraphics[height=4cm]{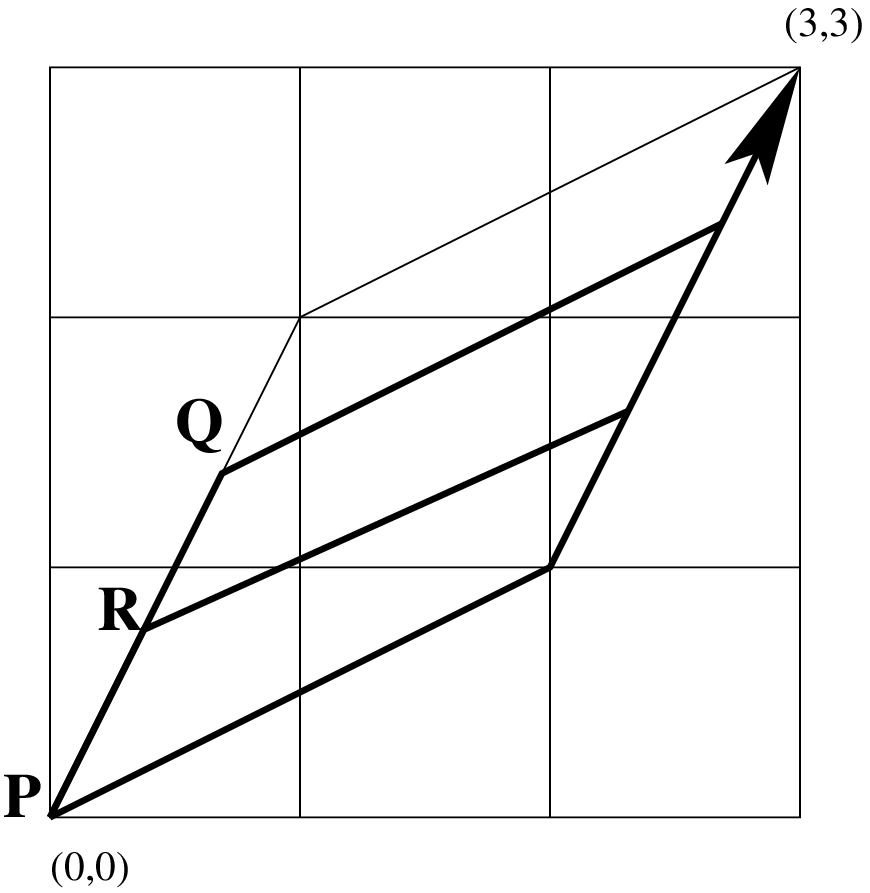}
\hspace{2cm}
\includegraphics[height=4cm]{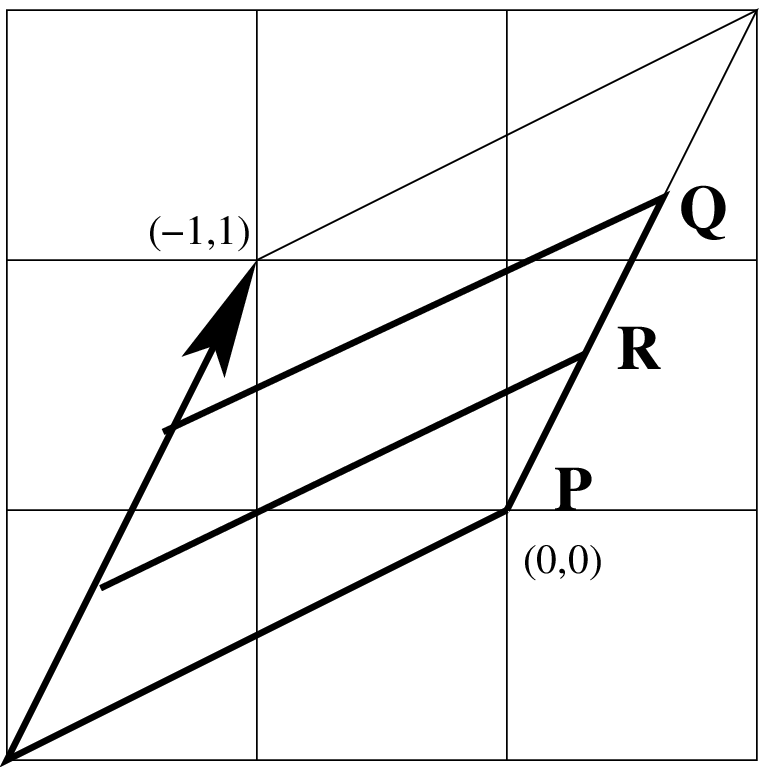}
\caption{ The reroutings $(1,2)S(2,1)$ and $(1,2)S(-2,-1)$ for $S=P,R,Q$ }
\label{p20a}
\end{figure}

A rerouted path is denoted $p_1Sp_2$, or $p_1Sp_2^{-1}$, and is the path that follows $p_1$ as far as the intersection point $S$, then follows $p_2$ (or the inverse loop $p_2^{-1}$) from $S$ back to $S$, and finally proceeds along $p_1$ from $S$ back to the starting point of $p_1$. This can be thought of as `inserting' the path $p_2$ into the path $p_1$ at the intersection point $S$. Note that the point $S$ may occur at the origin of $\IR^2$, in which case we follow $p_2$ straight away. An example of the intersections at the points $P$ (the origin), $R$ and $Q$ for the paths $p_1=(1,2)$ and $p_2=(2,1)$ (and its inverse $(-2,-1)$) is shown in Figure \ref{p20a}.  

\begin{figure}[hbtp]
\centering
\includegraphics[width=3cm]{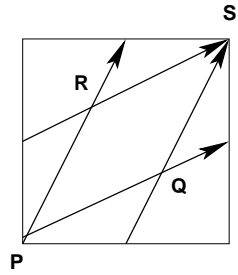}
\caption{The intersections $P,Q,R$ of the straight paths $(1,2)$ and $(2,1)$ }
\label{ga4}
\end{figure}

The concept of fundamental reduction described in Section \ref{sub21} can be used to simultaneously show the reduction of the paths $(1,2)$ and $(2,1)$ and, more importantly, their three intersections $P,Q,R$, as shown in Figure \ref{ga4}. Note that the point $S=(1,1)$ does not contribute since it coincides with the point $P=(0,0)$. Figure \ref{ga4} should be compared with Figure \ref{p20a}.

\begin{figure}
\begin{center}
\includegraphics[width=12pc]{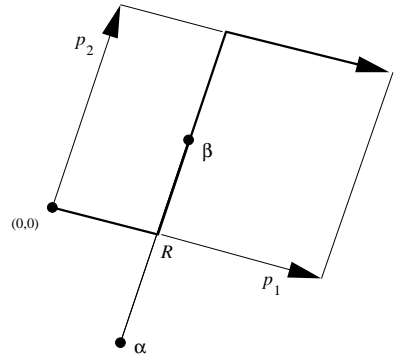}
\end{center}
\caption{\label{rerouting} The parallelogram with edges $p_1$ and $p_2$, and the rerouting $p_1Rp_2$ displayed as dark segments. The segment between integer points $\alpha$ and $\beta$ is a parallel copy of $p_2$, translated from starting at the origin to starting at $\alpha$; the grid of all other integer points is not displayed. }
\end{figure}

Here we shall work directly in $\IR^2$. Even though all integer points are the same when projected down to the torus a very clear picture of where the intersection point $R$ is located along both paths is obtained by fixing $p_1$ and parallel translating $p_2$ to start at a new integer point, denoted $\alpha$, in such a way that it intersects $p_1$ at $R$ (see Figure 
\ref{rerouting}). 

Let $(p)_{A,B}$ denote the subpath of the path $p$ from the point $A$ on $p$ to the point $B$ on $p$. Also, for any path $p$, let $\overline{p}$ denote the integer endpoint of $p$. For straight paths $p=(m,n)$, the notation for the endpoint and for the path itself coincide, i.e. $\overline{p}=p=(m,n)$,  but when the path is non-straight, there is a distinction between $p$ and 
$\overline{p}$. This latter situation occurs when we discuss the extension of the refined bracket (to be discussed in the next subsection) to `crooked' paths, i.e. non-straight paths, resulting from a former rerouting, like, for example, those of Figure \ref{p20a}.

Write $p_1= (p_1)_{OR}(p_1)_{R\overline{p_1}}$, to indicate how $p_1$ is divided into two segments by the intersection point $R$. Here $p_1$ starts at the origin $O$, but the parallel translated $p_2$ starts at the integer point $\alpha$. Let $p^\alpha$ denote the path $p$ parallel translated to start at $\alpha$ instead of its original starting point. (Of course, projected down to the torus, the original $p$ and the shifted $p^\alpha$ give rise to identical loops). In this way, the explicit algebraic expression for the rerouting $p_1Rp_2$ represented as a path in the plane (the four dark segments of Figure \ref{rerouting}) is:
\be
p_1Rp_2 = (p_1)_{O,R}\, (p_2^\alpha)_{R,\beta}\, (p_2^\beta)_{\beta, R+\overline{p_2}} \,
(p_1^{\overline{p_2}})_{R+ \overline{p_2},\overline{p_1} +\overline{p_2}}, 
\label{reroute1}
\ee
where the start and endpoints $\alpha$ and $\beta$ of $p_2^\alpha$ are related by: $\beta = 
\alpha + \overline{p_2}$. 

It is therefore clear why intersections occur and how they give rise to reroutings, i.e. for an intersection to occur, it is necessary that either $p_2^\alpha$ (the path $p_2$ parallel translated to start at $\alpha$) intersects $p_1$ in a point $R$ (which may be the origin, but not the endpoint of $p_1$), or equivalently, the endpoint $\beta$ is such that the appropriate $p_2^\alpha$ ending in $\beta$ intersects $p_1$ in a point $R$ as before.

\begin{figure}
\begin{center}
\includegraphics[width=12pc]{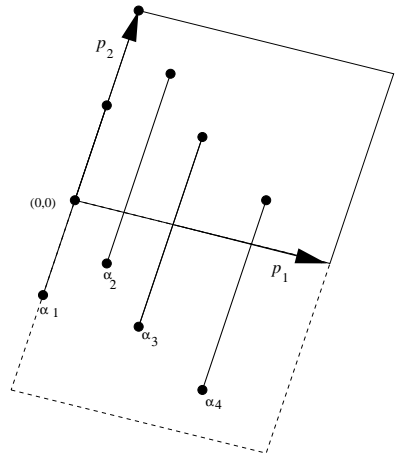}
\end{center}
\caption{\label{preparallelogram} The parallelogram and pre-parallelogram (dotted line) for $p_1$ and $p_2$, showing some of the integer points (black dots) and the corresponding parallel-translated copies of $p_2$}
\end{figure}

The possible starting points $\alpha$ are the integer points lying in a `pre - parallelogram' with vertices $-\overline{p_2}$, $-\overline{p_2} +\overline{p_1}$, $\overline{p_1}$ and the origin $O$. See Figure \ref{preparallelogram}. Here any integer points lying on the edge between $-\overline{p_2} +\overline{p_1}$ and $\overline{p_1}$ are excluded, since the corresponding paths $p_2^\alpha$ intersect $p_1$ at its endpoint. Similarly, any integer points lying on the edge between 
$-\overline{p_2}$ and $-\overline{p_2} +\overline{p_1}$ are excluded, to avoid double counting, since the integer points lying along the edge between $O$ and $\overline{p_1}$ are included. 

Equivalently, consider the endpoints $\beta$ lying inside the parallelogram generated by $p_1$ and $p_2$, i.e. with vertices $O$, $\overline{p_1}$, $\overline{p_1} + \overline{p_2}$ and $\overline{p_2}$. Here the integer points lying on the edge between  $\overline{p_1}$ and $\overline{p_1} +\overline{p_2}$ and those lying along the edge between $O$ and $\overline{p_1}$ are excluded (they correspond to the starting points we excluded from the pre-parallelogram).

From both perspectives it can be checked that the number of intersections is correct, since the total intersection number is: 
\be 
\epsilon (p_1,p_2) = \det \left(p_1 p_2\right).
\label{int}\ee
The total intersection number (counting multiplicities), all of whose contributions have the same sign, is therefore the modulus of $\epsilon (p_1,p_2)$, i.e. in geometric terms, the area of the parallelogram or the pre-parallelogram. In turn, this area is given by a classical theorem of Pick \cite{pick}, which states that the area $A(P)$ of a planar polygon $P$ with
vertices at integer points of the plane is given in terms of the number of
interior integer points $I(P)$ and the number of boundary integer points $B(P)$
as follows:
\be
A(P) = I(P) + \frac{B(P)}{2} -1.
\label{pick}\ee
This value is evidently the same as the number of integer starting points $\alpha$ in the pre-parallelogram, or equivalently the number of integer end points $\beta$ in the parallelogram generated by $p_1$ and $p_2$, since we have excluded two edges, i.e. half the integer points lying along the interior of the edges and three out of four of the vertices (this is accounted for by the $-1$ in the formula, since $\frac{4}{2} -1 =1$, the single remaining vertex). 

This approach clearly allows us to handle reducible paths in a natural manner. If $p_2$ is reducible, this means that there are integer points along $p_2$ other than the endpoints. In terms of the pre-parallelogram analysis, there will be $c$ integer starting points $\alpha$ for each intersection point $R$ along $p_1$. The intersection point $R$ therefore has intersection number $c$ and the rerouting at $R$ will appear with multiplicity $c$. These issues were briefly touched upon at the end of Section 5 of \cite{goldman}.

Since any rerouting must pass through one of the integer points inside the parallelogram generated by $p_1$ and $p_2$ (namely the appropriate endpoint $\beta$), we may label all intersections by the integer points that the rerouted paths (rerouted at each intersection) must pass through, rather than by their corresponding intersection points. 

\subsection{Two quantizations of the Goldman bracket}\label{sub32}
When analysing the behaviour of the Wilson loops $\hat{T}(p)= {\rm tr}~\hat{U}_{p}$ for the quantum connection \rref{conn} in \cite{goldman} a link with the Goldman bracket \cite{gol} emerged. This bracket is a Poisson bracket for the traces $T(\gamma)= {\rm tr}\, U_\ga$, defined on homotopy classes of loops on a surface, which for $U_\ga \in SL(2,\IR)$ takes the following form (see
\cite{gol} Thm. 3.14, 3.15 and Remark (2), p. 284):
\be
\{T(\ga_1), T(\ga_2)\} = \sum_{S \in \ga_1 \sharp \ga_2}
\epsilon(\ga_1,\ga_2,S)(T(\ga_1S\ga_2) -
T(\ga_1S\ga_2^{-1})).
\label{gold}
\ee
Here $\ga_1 \sharp \ga_2$ denotes the set of (transversal) intersection points
of $\ga_1$ and $\ga_2$, and $\epsilon(\ga_1,\ga_2,S)$ is the intersection number
of $\ga_1$ and $\ga_2$ at the intersection point $S$. One should note the rerouted paths $\ga_1S\ga_2$ and
$\ga_1S\ga_2^{-1}$ on the r.h.s. of \rref{gold}.

We found that the Wilson loops $\hat{T}(p)$, for $p=(m,n)$ a straight path corresponding to a loop on the torus, satisfied a quantum version of the Goldman relation \rref{gold}. In fact two different quantizations emerged, a direct one, and a refined one, which we will now describe.

First, for straight paths $p_1=(m,n)$ and $p_2=(s,t)$, the Goldman bracket takes the form:
\be 
\left\{T(m,n), T(s,t)\right\} = (mt-ns)
(T(m+s,n+t)- T(m-s,n-t)). 
\label{straightforward}
\ee 
Here $mt-ns$ is the total intersection number between $p_1$ and $p_2$, and this factor appears because there are effectively $mt-ns$ simple intersection points and the corresponding reroutings  $p_1Sp_2$ are all homotopic to the straight path $(m+s, n+t)$, with an analogous statement for the negative reroutings $p_1Sp_2^{-1}$. 

By a simple direct calculation, it was shown in \cite{goldman} that the Wilson loops satisfy:
\begin{equation}
[\hat{T}(m,n), \hat{T}(s,t)]=
(q^{\frac{mt-ns}{2}}-q^{-\frac{mt-ns}{2}}) \left(\hat{T}(m+s,n+t) - \hat{T}(m-s,n-t)\right)
\label{qgb1}
\end{equation}
i.e. a quantization of \rref{straightforward}, with the total intersection number $mt-ns$ replaced by a quantum total intersection number (the first factor on the r.h.s. of \rref{qgb1}). We call this the direct quantization of the Goldman bracket (previously referred to as preliminary, or straightforward, or unrefined).

A refined quantization was also obtained in \cite{goldman}, where each rerouting appears as a separate term, and the relative area phases of these different but homotopic reroutings are taken into acccount. This takes the following form:
\be
[\hat{T}(p_1), \hat{T}(p_2)] = \sum_{ {S} \in p_1 \sharp p_2}
(q^{\epsilon(p_1,p_2,{S})} - 1)\hat{T}(p_1{S}p_2)  
+ (q^{-\epsilon(p_1,p_2,{S})} - 1)\hat{T}(p_1{S}p_2^{-1})
\label{qgold}
\ee
which quantizes the bracket \rref{gold} (with loops $\ga$ substituted by paths $p$) by replacing the intersection numbers $\epsilon(p_1,p_2,{S})$ by quantum intersection numbers $(q^{\epsilon(p_1,p_2,{S})} - 1)$. 

An example of equation \rref{qgold} is given by the choice $p_1=(1,2)$, $p_2=(2,1)$
\be
[\hat{T}(1,2),\hat{T}(2,1)] =\sum_{S=P,R,Q} 
(q^{-1} -1) \hat{T}((1,2)S(2,1)) + (q-1) \hat{T}((1,2)S(-2,-1)), 
\label{qgoldexp}
\ee
where the terms on the r.h.s. come from the positive and negative reroutings at three intersection points denoted $P,\,R,\,Q$, see Figure \ref{p20a}. To relate \rref{qgoldexp} to the direct form \rref{qgb1} of this commutator we calculate the area phases of each rerouting on the r.h.s. of 
\rref{qgoldexp} relative to the corresponding straight paths $(3,3)$ or $(-1,1)$, and using 
\begin{equation}
\hat{T}({p_1})=q^{S(p_1,p_2)}\hat{T}({p_2}),
\label{areaphasewilson}
\end{equation}
which follows from \rref{areaphase} we obtain:
\begin{eqnarray}
[\hat{T}(1,2),\hat{T}(2,1)] &=& (q^{-1}-1)(q^{\frac{3}{2}} + q^{\frac{1}{2}} + q^{-\frac{1}{2}}) \hat{T}(3,3)\nonumber \\
&+& (q-1) (q^{-\frac{3}{2}} + q^{-\frac{1}{2}} + q^{\frac{1}{2}}) \hat{T}(-1,1) \nonumber \\
&=& (q^{-\frac{3}{2}} - q^{\frac{3}{2}}) (\hat{T}(3,3) - \hat{T}(-1,1)).
\end{eqnarray}
as required. 

\section{Properties of commutators of Wilson loops}\label{sec4}
To obtain a fully consistent theory for intersecting loops, based on the refined bracket \rref{qgold}, we need to address the following issue: the paths on the l.h.s. are straight, whereas on the r.h.s. they are crooked, so for the bracket to close, we need to understand the refined bracket with crooked paths on the l.h.s.

When attempting to extend the refined bracket to crooked paths, we encountered unexpected extra phases when the second path $p_2$ in the rerouting $p_1Rp_2$ was crooked, and the first path $p_1$ was straight. These difficulties did not occur when the first path was crooked and the second straight, i.e. the discrepancy occurs when the middle part of the rerouting (the inserted path) is crooked. To better understand these `crooked reroutings' and the extra phases they generate, we derive a formula for the relative phase of such reroutings compared to the `first' rerouting, i.e. the rerouting that occurs straight away at the origin (the origin is always an intersection point for any two paths).

\subsection{The relative phase for a crooked rerouting}\label{sub41}
\begin{figure}
\begin{center}
\includegraphics[width=10pc]{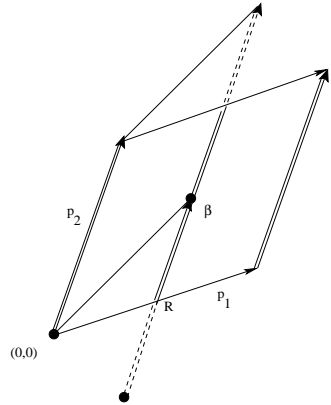}
\end{center}
\caption{\label{crookedrerouting} Rerouting along a crooked path $p_2$, represented as a double line.}
\end{figure}

We refer to Figure \ref{crookedrerouting} for the following discussion and calculations. There we have written $p_2$ as a double arrow, to indicate that it is crooked, i.e. it comes from one or more previous rerouting processes. The endpoint of the double arrow is the endpoint of $p_2$, written $\overline{p_2}$ using the notation of Section \ref{sub31}. 
We derive an expression for the signed area, or relative phase, between the rerouting at the intersection point $R$, namely $p_1Rp_2$ and the rerouting at the origin (i.e. the first rerouting that occurs when starting along $p_1$).

The `crooked rerouting' may be expressed exactly like \rref{reroute1} as:
\be
p_1Rp_2 = (p_1)_{O,R}\, (p_2^\alpha)_{R,\beta}\, (p_2^\beta)_{\beta, R+\overline{p_2}} \,
(p_1^{\overline{p_2}})_{R+ \overline{p_2},\overline{p_1} +\overline{p_2}}, 
\label{reroute2}\ee
where the start and endpoints $\alpha$ and $\beta$ of $p_2^\alpha$ are related by: $\beta = 
\alpha + \overline{p_2}$, whereas the rerouting at the origin is:
\be
 p_1Op_2 = (p_2)_{O, \overline{p_2}}\, (p_1^{\overline{p_2}})_{\overline{p_2}, R+ \overline{p_2}}
\label{reroute3}\ee
This is all that is needed since the relative phase of $p_1Op_2$ compared to the natural reference path, i.e. the straight path from the origin to $\overline{p_1}+ \overline{p_2}$ is easily calculated.  

The signed area $S(p_1Rp_2, p_1Op_2)$ between \rref{reroute2} and \rref{reroute3} is found in two steps, by geometric adjustments of the area we are considering. First, notice that we can omit the last part of the two paths $(p_1^{\overline{p_2}})_{R+ \overline{p_2},\overline{p_1} +\overline{p_2}}$, which is the same for both reroutings, and can add and subtract a `triangle' with two straight edges, $(p_1)_{O,R}$ and $(\beta)_{O,\beta}$ (the path going from the origin to the integer point $\beta$), and one crooked edge $(p_2^\alpha)_{R,\beta}$. This gives
\be
S(p_1Rp_2, p_1Op_2) = S( (\beta)_{O,\beta}\, (p_2^\beta)_{\beta, \beta+\overline{p_2}},
(p_2)_{O, \overline{p_2}} (\beta)_{\overline{p_2},\beta + \overline{p_2}}).
\label{s1}\ee
Then we simply replace the two parallel crooked edges of the parallelogram, i.e. $(p_2^\beta)_{\beta, \beta+\overline{p_2}}$ and $(p_2)_{O, \overline{p_2}}$, by the corresponding straight edges, to obtain:
\bea
 && S( (\beta)_{O,\beta}\, (p_2^\beta)_{\beta, \beta+\overline{p_2}},
(p_2)_{O, \overline{p_2}} (\beta)_{\overline{p_2},\beta + \overline{p_2}})\nonumber\\
 && = S( (\beta)_{O,\beta}\, (\overline{p_2}^\beta)_{\beta, \beta+\overline{p_2}},
(\overline{p_2})_{O, \overline{p_2}} (\beta)_{\overline{p_2},\beta + \overline{p_2}}) =
 \det (\beta \overline{p_2}).
\label{s2}\eea
That is, the signed area, or relative phase, between the rerouting $p_1Rp_2$ and the rerouting at the origin $p_1Op_2$ is
\be
 S(p_1Rp_2, p_1Op_2) = \det (\beta \overline{p_2})
\label{s3}\ee
where $\beta$ is the integer endpoint associated to the intersection point $R$. This 
result will be used in Section \ref{sub43}. 

\subsection{A geometric formula}\label{sub42}
Here we use the classification by integer points of Section \ref{sub31} to obtain a quantum version of equation 
\rref{pick}, namely Pick's formula \cite{pick}.
 
We define a set $V$ of valid integer points associated to the parallelogram $P(p_1,p_2)$ generated by the straight paths $p_1$ and $p_2$. This set consists of all interior integer points, all non-vertex integer points lying along $p_2$ and $p_1^{\overline{p_2}}$, as well as the vertex integer point $p_2$ itself. The significance of the points of $V$ in relation to the previous sections is that these are endpoints of parallel copies of $p_2$ with starting points in the pre-parallelogram of $P(p_1,p_2)$.

The number of points $\bf V$, in $V$ is equal to the area $A(P)$ of $P(p_1,p_2)$ because of Pick's theorem \rref{pick}: 
\be
{\bf V} =  \sum_{\beta \in V} 1 = I(P) + B(P)/2 -1 = A(P).
\label{pick2}\ee
This is clear since the number of non-vertex integer points lying along $p_2$ and $p_1^{\overline{p_2}}$ is half the total number of non-vertex integer points on the boundary, i.e.  $(B(P) -4)/2= B(P)/2-2$. Adding the single vertex point $p_2$ gives the r.h.s. of Pick's formula.
  
\begin{figure}
\begin{center}
\includegraphics[width=25pc]{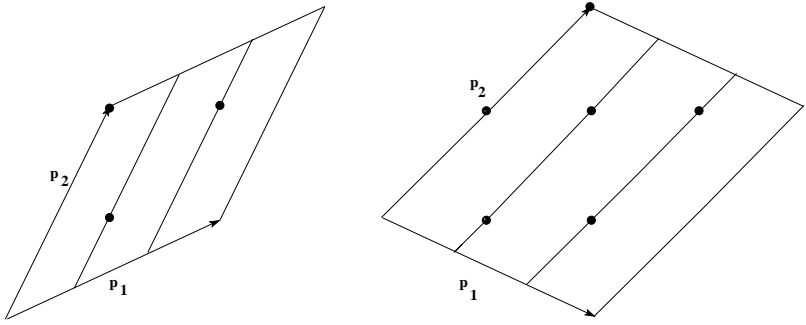}
\end{center}
\caption{\label{geomformulafig} Elements of V (black dots) for $p_2$ irreducible (left) and reducible (right).}
\end{figure}

Consider first the case where $p_1$ and $p_2$ have a positive intersection number $\det(p_1,p_2)>0$, and $p_2$ is irreducible (see Figure \ref{geomformulafig}). We have the following formula:
\be
\sum_{\beta \in V}q^{\det(\beta,p_2)} = \frac{q^{\det(p_1,p_2)}-1}{q-1}
\label{geomformula}
\ee
To see this, note that any line parallel to $p_2$ in the parallelogram has, since $p_2$ is irreducible, at most one point of $V$ lying in it. It follows that there are precisely $N=\det(p_1,p_2)$ such lines passing through a point of $V$ (including $p_2$ itself), and these lines divide the parallelogram into $N$ thin parallelograms, each of area $1$ (see Figure \ref{geomformulafig}). The area of all the thin parallelograms between the point $\beta$ and the edge $p_2$ is equal to $\det(\beta,p_2)$. Therefore
\be
\sum_{\beta \in V}q^{\det(\beta,p_2)}=\sum_{i=0}^{N-1}q^{i}= \frac{q^{N}-1}{q-1}.
\label{det}\ee

For the case $\det(p_1,p_2)<0$, and $p_2$ irreducible, the geometric formula \rref{geomformula} is modified (equivalent to replacing $q$ with $q^{-1}$) to:
\be
\sum_{\beta \in V}q^{\det(\beta,p_2)} = \frac{q^{\det(p_1,p_2)}-1}{q^{-1}-1}
\label{geomformula_negint}
\ee
This follows by analogous arguments starting from the geometric series formula \rref{geomformula}
with $q$ replaced by $q^{-1}$.

Suppose now that again $p_1$ and $p_2$ have a positive intersection number, i.e. 
$\det(p_1,p_2)>0$, but that $p_2$ is reducible, in other words $p_2$ has $M>1$ points of $V$ lying along it. The area of $P(p_1,p_2)$ is an integer multiple of $M$, say $\det(p_1,p_2)=NM$, and the $NM$ points of $V$ lie along $N$ lines parallel to $p_2$ (see Figure \ref{geomformulafig}), with $M$ points in each line. These lines divide the parallelogram into $N$ thin parallelograms, each of area $M$. We have a new geometric formula:
\be
\frac{1}{M}\sum_{\beta \in V}q^{\det(\beta,p_2)} = \frac{q^{\det(p_1,p_2)}-1}{q^M-1}
\label{geomformula_red}
\ee
which follows by grouping together the $M$ equal terms coming from points $\beta$ lying on the same parallel line and using the geometric series formula:
 $$
 \sum_{i=0}^{N-1}q^{Mi}= \frac{q^{NM}-1}{q^{M}-1}
$$

Finally the analogous formula for the case $\det(p_1,p_2)<0$ and $p_2$ reducible is as follows:
\be
\frac{1}{M}\sum_{\beta \in V}q^{\det(\beta,p_2)} = \frac{q^{\det(p_1,p_2)}-1}{q^{-M}-1}
\label{geomformula_red_negint}
\ee

We remark that the geometric formula \rref{geomformula} can be interpreted as a quantum version of \rref{pick2}. That is, the r.h.s. of \rref{geomformula} is the quantum area of the parallelogram $P(p_1,p_2)$, is expressed as a sum over terms labelled by $\beta \in V$, each of which equals $1$ in the classical limit $q\rightarrow 1$.  A simple example is given by $p_1=(2,1)$ and $p_2=(1,2)$,  with a corresponding quantum area of $1 + q + q^2$, instead of, classically, 3. We conjecture that this interpretation can be generalised in some way to an arbitrary polygon $P$.

\subsection{Equivalence of the refined and direct brackets for straight paths}\label{sub43}
We now show, for the general case, how the direct and refined formulas \rref{straightforward} and \rref{qgold} for $[\hat{T}(p_1),\hat{T}(p_2)]$ are equivalent, when $p_1$ and $p_2$ are
straight.  When $p_2$ is irreducible and $\det(p_1,p_2)>0$, the refined formula \rref{qgold} gives:
\begin{eqnarray*}
[\hat{T}(p_1), \hat{T}(p_2)]& = &(q-1) \sum_{\beta \in V(p_1,p_2)}
q^{\det(\beta,p_2)} \hat{T}(p_2 p_1^{\overline{p_2}})\\& + &
 (q^{-1}-1) \sum_{\gamma \in V(p_1,p_2^{-1}) }
q^{\det(\gamma,p_2^{-1})} \hat{T}(p_2^{-1} p_1^{-\overline{p_2}})
\end{eqnarray*}
where we have used \rref{s3} to obtain the geometric factors
of the reroutings associated to an integer point $\beta$ (or $\gamma$)
relative to the respective `first reroutings' (i.e. reroutings at
the origin). With the geometric formulae  \rref{geomformula} and
\rref{geomformula_negint} we obtain:
\begin{eqnarray}
[\hat{T}(p_1), \hat{T}(p_2)] & = & (q^{\det(p_1,p_2)}-1)  \hat{T}(p_2
p_1^{\overline{p_2}}) +  (q^{\det(p_1,p_2^{-1})}-1) \hat{T}(p_2^{-1} p_1^{-\overline{p_2}})\nonumber \\
& = & (q^{\det(p_1,p_2)}-1) q^{-\det(p_1,p_2)/2} \hat{T}(p_1+p_2) + \nonumber \\
& & \quad (q^{\det(p_1,p_2^{-1})}-1) q^{-\det(p_1,p_2^{-1})/2} \hat{T}(p_1-p_2) \nonumber \\
& = & (q^{\det(p_1,p_2)/2}- q^{-\det(p_1,p_2)/2}) (\hat{T}(p_1+p_2) - \hat{T}(p_1-p_2))\nonumber \\
\label{refined-unref}
\end{eqnarray}
i.e. the direct formula  \rref{straightforward}, where we have used
the relative geometric factors for the `first' reroutings relative
to the corresponding straight diagonal paths $p_1 \pm p_2$, as well as
the property $\det(p_1, p_2^{-1}) = -\det(p_1,p_2)$.

As discussed briefly in \cite{goldman}, when $p_2$ is
reducible, the refined formula must be modified by the introduction of
quantum {\em multiple} intersection numbers. Again let
$\det(p_1,p_2)>0$, and suppose $p_2$ has $M>1$ points of $V$ along it. The intersection number of the intersection corresponding to $M$ integer points lying on a line in $P(p_1,p_2)$ parallel to $p_2$, is $M$ classically, and the quantum multiple intersection number is $q^M-1$. The refined formula for $[\hat{T}(p_1), \hat{T}(p_2)]$ when $p_2$ is reducible of multiplicity $M$, is given by:
\be
[\hat{T}(p_1), \hat{T}(p_2)] = \sum_{ {S} \in p_1 \sharp p_2}
(q^{M} - 1)\hat{T}(p_1{S}p_2)
+ (q^{-M} - 1)\hat{T}(p_1{S}p_2^{-1})
\ee
Each intersection point $S$ is associated to $M$ integer points, so this can be rewritten as 
\begin{eqnarray*}
[\hat{T}(p_1), \hat{T}(p_2)] &=& \frac{(q^{M} - 1)}{M}
 \sum_{ \beta \in V(p_1,p_2)} q^{\det(\beta,p_2)} \hat{T}(p_2 p_1^{\overline{p_2}}) \\  
&+&\frac{(q^{-M} - 1)}{M} \sum_{ \gamma \in V(p_1,p_2^{-1})} q^{\det(\gamma,p_2^{-1})}
\hat{T}(p_2^{-1} p_1^{-\overline{p_2}})
\end{eqnarray*}
(recall that the factors $q^{\det(\beta,p_2)}$ are the same for all points $\beta$ lying on a line parallel to $p_2$). 

Now, using the corresponding geometrical formula \rref{geomformula_red}, we obtain:
\begin{eqnarray*}
[\hat{T}(p_1), \hat{T}(p_2)]& = & (q^{\det(p_1,p_2)}-1)
\hat{T}(p_2 p_1^{\overline{p_2}}) +  (q^{-\det(p_1,p_2)}-1)
\hat{T}(p_2^{-1} p_1^{-\overline{p_2}}) \\
& = & (q^{\det(p_1,p_2)/2}- q^{-\det(p_1,p_2)/2}) (\hat{T}(p_1+p_2) - \hat{T}(p_1-p_2)
\end{eqnarray*}
where the steps in the final equality are the same as those in \rref{refined-unref}.

Finally the cases where $\det(p_1,p_2)<0$ are dealt with by using the geometric formulae \rref{geomformula_negint} and \rref{geomformula_red_negint} respectively, since the terms associated with a positive rerouting $p_1Sp_2$ now appear with a factor $q^{-1}-1$, and the terms associated with a negative rerouting $p_1Sp_2^{-1}$ appear with a factor $q-1$. In all cases the direct form \rref{straightforward} of the bracket $[\hat{T}(p_1), \hat{T}(p_2)]$ is found. 

\subsection{Antisymmetry of the refined bracket}\label{sub44}
\begin{figure}
\begin{center}
\includegraphics[width=10pc]{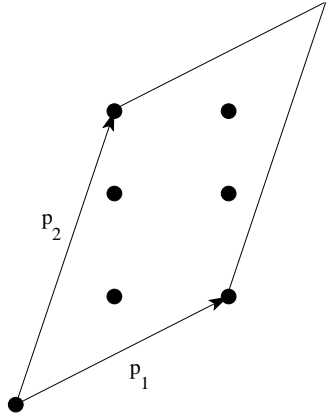}
\end{center}
\caption{\label{antisymmetry} V($p_1,p_2$) contains the four internal points and $p_2$, V($p_2,p_1$) contains the four internal points and $p_1$.}
\end{figure}

Finally, we show how the analysis of the previous section permits the antisymmetry of the refined bracket 
\be
[\hat{T}(p_1), \hat{T}(p_2)] = - [\hat{T}(p_2), \hat{T}(p_1)]
\label{antisymm}
\ee
to be deduced directly from its expression \rref{qgold}, rather than indirectly from the equality of the refined and direct bracket, together with the patent antisymmetry of the latter. Instead of a general argument, we will focus on the particular case where $p_1$ and $p_2$ are both irreducible, $\det(p_1,p_2)>0$, and will only consider the positive rerouting terms, illustrating the arguments with the example of Figure \ref{antisymmetry} where $p_1=(2,1)$ and $p_2=(1,3)$.

The contribution of the positive rerouting terms to (\ref{antisymm}), using the results from the previous section, is:
$$
(q-1) \sum_{\beta \in V(p_1,p_2) } q^{\det(\beta,p_2)} \hat{T}(p_2 p_1^{\overline{p_2}}) =
- (q^{-1}-1) \sum_{\beta' \in V(p_2,p_1) } q^{\det(\beta',p_1)} \hat{T}(p_1 p_2^{\overline{p_1}}).
$$ 
By factoring out  $q^{-1}$ on the r.h.s., and using 
$$
\hat{T}(p_1 p_2^{\overline{p_1}}) = q^{\det(p_1,p_2)}  \hat{T}(p_2 p_1^{\overline{p_2}}),
$$
it remains to prove the equality:
\be
 \sum_{ \beta \in V(p_1,p_2)} q^{\det(\beta,p_2)} = q^{-1} q^{\det(p_1,p_2)} \sum_{ \beta'' \in V(p_2,p_1)} q^{\det(\beta',p_1)}
\label{sumformula}
\ee
Note that the sets $V(p_1,p_2)$ and $V(p_2,p_1)$ share the same internal integer points, and differ only in that $V(p_1,p_2)$ contains $p_2$ and not $p_1$, whereas $V(p_2,p_1)$ contains $p_1$ and not $p_2$. The exponents $\det(\beta',p_1)$ on the r.h.s. are disjoint non-positive integers, running from $0$ to $-\det(p_1,p_2) + 1$, and represent the negative of the areas of the parallelograms bounded on the left and right by $p_2$ and $p_2^{\overline{p_1}}$, bounded below by $p_1$ and above by a line parallel to $p_1$ through $\beta'$.  Multiplying the sum on the r.h.s. by $q^{-1}$ corresponds to shifting the range of negative areas to run from $-1$ to $-\det(p_1,p_2)$, and taking the complementary areas $-\det(p_1,p_2)| -(\det(\beta',p_1)-1)$ (corresponding to multiplying on the r.h.s. with $q^{\det(p_1,p_2)}$), which are positive, these will run from $0$ to $\det(p_1,p_2)-1$. These latter exponents are exactly the exponents $\det(\beta,p_2)$ that appear on the l.h.s. of (\ref{sumformula}), i.e. corresponding to the (positive) areas of parallelograms bounded above and below by $p_1$ and $p_1^{\overline{p_2}}$, on the left by $p_2$ and on the right by the line parallel to $p_2$ through $\beta$.

This geometric argument establishes the equality (\ref{sumformula}) and hence the antisymmetry (\ref{antisymm}) for the case considered.

\section{Conclusions}\label{concl}
We have argued that this model of $2+1$ quantum gravity implies considering Wilson variables 
for a large class of loops, related by area phases. Our study has substantially clarified the nature of the Wilson observables for crooked paths, namely paths that come from intersecting straight paths and rerouting along one of them. It has prepared us for a fuller description of the refined Goldman bracket for this larger class of loops, i.e. the class that contains loops coming from both straight paths and crooked paths. 

In particular, 
\begin{itemize}
\item we have given a full proof showing that, for straight paths, the refined bracket and the direct bracket are equivalent.
\item This proof also deals satisfactorily with reducible paths, i.e. paths such that $p=(m,n)=c(m',n')$, where $m,n,c,m',n'$ are all integers, with $c\geq 2$.
\item Our methods are a substantial step forwards towards our ultimate goal of defining the refined bracket which closes on a suitable class of straight and crooked paths.
\item We have achieved an independent understanding of the antisymmetry of the refined bracket for two straight paths, which is not manifest.
\end{itemize}

Apart from the main goal mentioned above, there are a couple of other questions which should be addressed in a fully consistent intersection and rerouting theory based on the refined bracket: 

\begin{itemize}
\item The quantum intersection numbers which naturally appear in the two quantizations \rref{qgb1} and \rref{qgold} have a different appearance, and in particular they are symmetric, for the direct bracket, under the interchange $q \leftrightarrow q^{-1}$. This is not the case for the refined bracket.

\item We must prove the Jacobi identity for the refined bracket extended to a wider class of paths.
\end{itemize}

In future work we hope to explore in more detail the link with $2$-dimensional holonomy \cite{FMP}, as discussed at the end of Section 2, and gain further understanding of the elegant quantum geometry that emerges through the quantized Goldman bracket, relating it e.g. to noncommutative geometry \cite{con}, or to other quantizations of the Goldman bracket \cite{Turaev}, or the BTZ black hole \cite{vazwit}.

\begin{acknowledgements}
This work was supported by the Istituto Nazionale di Fisica Nucleare (INFN) of Italy, Iniziativa Specifica FI41, the Italian Ministero dell' Universit\`a e della Ricerca Scientifica e Tecnologica (MIUR), and the projects Quantum Topology, POCI/MAT /60352/2004, and New Geometry and Topology, PTDC/MAT/101503/2008, financed by the {\em Funda\c{c}\~{a}o para a Ci\^{e}ncia e a Tecnologia} (FCT) and cofinanced by the European Community fund FEDER.
\end{acknowledgements}



\end{document}